\documentclass[9pt,conference]{IEEEtran}
\usepackage{amssymb,amsthm,amsmath,array}
\usepackage{graphicx}
\usepackage[caption=false,font=footnotesize]{subfig}
\usepackage{xspace}
\usepackage[sort&compress, numbers]{natbib}
\usepackage{stmaryrd}
\usepackage{xcolor}
\usepackage{mathtools}
\usepackage{float}
\usepackage{textcomp}
\usepackage{comment}
\usepackage{svg}

\newcommand{\bs}[1]{\boldsymbol{#1}}

\begin{document}
\title{Compressive spectral imaging\\ in view of Earth observation applications }
\author{\IEEEauthorblockN{
        Clément Thomas\IEEEauthorrefmark{1}, 
        Laurent Jacques\IEEEauthorrefmark{2}, and 
        Marc Georges\IEEEauthorrefmark{1}
    }
    \IEEEauthorblockA{
        \IEEEauthorrefmark{1} Centre Spatial de Liège (CSL), Université de Liège, Avenue du Pré-Aily, 4031 Liège, Belgium.\\
        \IEEEauthorrefmark{2} ICTEAM institute, UCLouvain, Avenue Georges Lemaître 4-6/L4.05.01, 1348 Louvain-la-Neuve, Belgium.\\}
}
\maketitle
\begin{abstract}
    Earth observation from space is an important scientific and industrial activity that has applications in many sectors. The instruments employed are often large, complex, and expensive. In addition, they generate large amounts of data, which is challenging for storage and transfer purposes. Compressive spectral imaging would be a cheaper, more efficient, and well-adapted technique to perform Earth observation. An interesting architecture is compressive spectral imaging with diffractive lenses, which is extremely compact. This work investigates the possibility of replacing the diffractive lens in this system with a classical refractive lens. Taking advantage of the chromatic aberration of a lens makes the use of expensive diffractive lenses unnecessary. Simulations are performed to test the feasibility of the method. Signal recovery is a basis pursuit solved using the Douglas-Rashford algorithm. 
\end{abstract}

\section{Introduction}
\begin{figure}[h]
    \centering
    \includegraphics[width=0.9\linewidth]{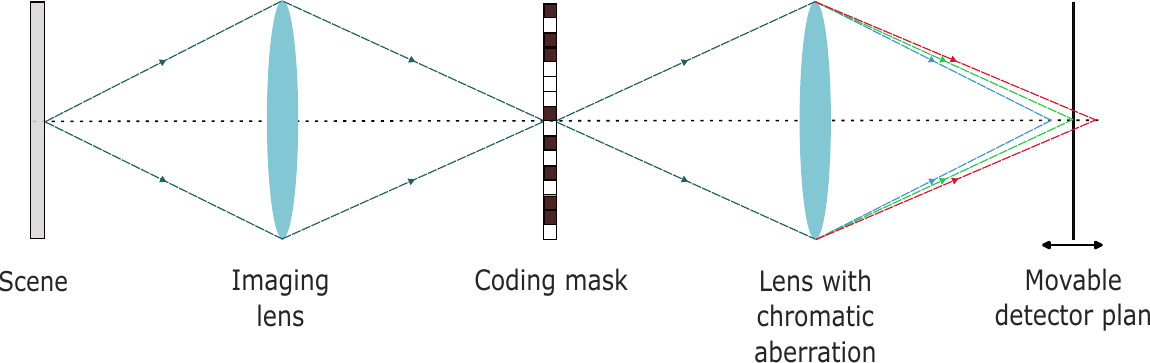}
    \caption{Proposed architecture.}
    \label{CSID}
\end{figure}

Satellite Earth observation plays a key role in modern society, having applications in meteorology, climatology, navigation, agriculture, or emergency management, to cite a few. Acquiring data for these purposes demands high spatial and spectral resolution imagers. This results in the design of large, complex, and costly instruments accompanied by the production of large amounts of data that pose challenges for both storage and transmission. Therefore, designing an instrument based on Compressive Sensing (CS) principles would be a cheaper and more efficient manner to perform Earth observation. Indeed, CS is a technique that reconstructs compressible signals from fewer measurements than traditional sampling methods require. In this idea, compressive spectral imaging takes advantage of the CS paradigm to sense 3D spectral images from a limited number of projections on a 2D detector. Successfully applying this technique in an Earth observation instrument would allow to get rid of the complex scanning systems of traditional push-broom or whisk-broom imagers and increase the acquisition time on each pixel, all while reducing data quantities.\\
The best known architecture is the Coded Aperture Snapshot Spectral Imaging (CASSI) system that has been developed in several variations \cite{gehm2007single, wagadarikar2008single, lin2014spatial, correa2015snapshot}. In the most general case, the spectral field is coded by a Spatial Light Modulator (SLM) and then passes through a dispersive element such as a prism or a diffraction grating to separate the wavelengths. The resulting light is then focused on a 2D monochrome detector. Individual retrieval of multiplexed spectral bands is enabled by leveraging precise knowledge of the coding and dispersing operations applied to the spectral field within a sparse recovery algorithm.
More recently, a CASSI-derived architecture, called Compressive Spectral Imaging with Diffractive lenses (CSID), was introduced in Ref. \cite{kar2019compressive}. In this system, the dispersive element is not a prism or a diffraction grating but a diffractive lens. The latter uses diffraction through microstructures to focus light at a distance varying with the wavelength. Hence, this element presents focusing and dispersive abilities simultaneously and replaces both the dispersive element and its related relay lens in a traditional CASSI system, further compacting the architecture.\\
This work investigates the possibility of replacing the diffractive lens with a classical refractive lens, which would make the system simpler and cheaper due to the complexity and cost of manufacturing diffractive lenses. In this case, dispersion can be achieved by selecting or designing a lens with sufficient chromatic aberration, i.e. a focal length that shifts with the wavelength. The proposed architecture is illustrated in Fig. \ref{CSID}. It must be noted that while the architecture is chosen to be suitable for Earth observation, it is adapted to any applications requiring fast, compressive, and compact spectral imagers.

\section{System model} 

\subsection{Forward model}
In this CSID-like configuration (Fig. \ref{CSID}), the source is first focused by a lens on a SLM that spatially modulates the signal. After that, the resulting signal passes through a dispersive lens, which focuses and disperses light simultaneously. The signal is then integrated by a monochromatic detector. The lens has a wavelength-dependent focal length, thus, a measurement is a superposition of differently blurred and coded spectral bands. To achieve measurement diversity, a total of $K$ such measurements can be recorded by changing the focus, which is done by slightly moving the detector along the optical axis.\\
The measurements obtained with this system can be written in a discretized version as
\begin{equation}
    y_k[m,n] = \sum_{s=1}^{S} \left( (\bs{x_s} \odot \bs{c_k}) \ast \bs{h_{s,k}} \right) [m,n],
\end{equation}
with the Hadamard product $(\bs{u} \odot \bs{v})[m,n] := u[m,n]v[m,n]$, and $\bs{u} \ast \bs{v}$ the spatial (periodic) convolution of $\bs{u}$ by $\bs{v}$, for any $N_x \times N_y$ images $\bs{u}$ and $\bs{v}$. $[m,n]$ represent the detector pixel coordinates, $\bs{y}_k$ is the $k$th measurement intensity, $S$ is the number of spectral bands, $\bs{x}_s$ is the $s$th band of the spectral field, $\bs{c}_k$ is the coded mask, and $\bs{h}_{s,k}$ is the Point Spread Function (PSF) of the $s$th spectral band at the $k$th measurement.
Finally, the discrete model can be expressed by
\begin{equation}
    \bs{y} = \bs{HCx} + \bs{n}
\end{equation}
where $\bs{y} = [\bs{y}_1^T,...,\bs{y}_K^T]^T \in \mathbb{R}^{KN}$ is the vertically concatenated measurement vector with $N = N_xN_y$ the number of detector pixels, $\bs{x} = [\bs{x}_1^T,...,\bs{x}_S^T]^T \in \mathbb{R}^{SN}$ is the concatenated image vector. The $KN \times SN$ matrix $\bs{H}$ consists of $N \times N $ convolution matrices representing the convolutions with PSFs $h_{s,k}$. More precisely, the $N \times N$ matrix located in the $k$th block row and $s$th block column of $\bs{H}$ is a block circulant matrix with circular blocks corresponding to the circular convolution operation with the PSF $h_{s,k}$. The diagonal matrix $\bs{C} \in \mathbb{R}^{N\times N}$ performs the coding operations and has values 0 or 1 along its diagonal. Finally, $\bs{n} = [\bs{n}_1^T,...,\bs{n}_K^T]^T$ represents the noise. Naturally, for the system to be compressive, the number of measurements $K$ must be smaller than the number of spectral bands $S$.

\subsection{Reconstruction}

The chosen reconstruction method is a Basis Pursuit (BP) \cite{chen2001atomic}, which may be written as
\begin{equation}
    \bs{\alpha^*} = \underset{\bs{\alpha}}{\text{arg min}}  \|\bs{\alpha}\|_1 \text{ s.t. }  \|\bs{y} - \bs{HC\Psi \alpha}\|_2 \leq \epsilon, 
    \label{inv_l1}
\end{equation}
where the reconstructed volume is $\bs{x^*} = \bs{\Psi \alpha^*}$, and $\epsilon$ is a scalar representing the noise tolerance. The sparsifying basis is chosen as the Kronecker product $\bs{\Psi} = \bs{\Psi_2} \otimes \bs{\Psi_1}^\intercal$ with $\bs{\Psi_1}$ a 2D discrete wavelet transform applied in the spatial dimensions and $\bs{\Psi_2}$ a 1D discrete cosine transform applied in the spectral dimension.\\

Signal recovery is done using the Douglas-Rashford (DR) algorithm, which is an iterative scheme to minimize the sum of closed proper convex functions that are not necessarily smooth \cite{combettes2011proximal} \cite{parikh2014proximal}. It was chosen for its accessible implementation and proven effectiveness, even though more complex methods using neural networks could achieve higher accuracy. In this application, it takes the form
\begin{equation}
    \underset{\bs{\alpha}}{\text{min}} f(\bs{\alpha}) + g(\bs{\alpha})
\end{equation}
with $g(\bs{\alpha}) = \|\bs{\alpha} \|_1$ and
\begin{equation}
    f(\bs{\alpha}) = 
    \begin{cases}
        0 & \text{ if } \bs{\alpha} \in \Omega, \\
        + \infty & \text{ if } \bs{\alpha} \notin \Omega,
    \end{cases}
\end{equation}
where $\Omega = \{\bs{\alpha} \in \mathbb{R}^{SN}, \| \bs{y} - \bs{A \alpha} \|_2 \leq \epsilon \}$ is an affine space and $\bs{A} = \bs{HC\Psi}$. It must be noted that at the moment, only noiseless cases ($\epsilon = 0$) have been simulated.

\section{Simulations}
\begin{figure}[b]
    \centering
    \includegraphics[width=0.9\linewidth]{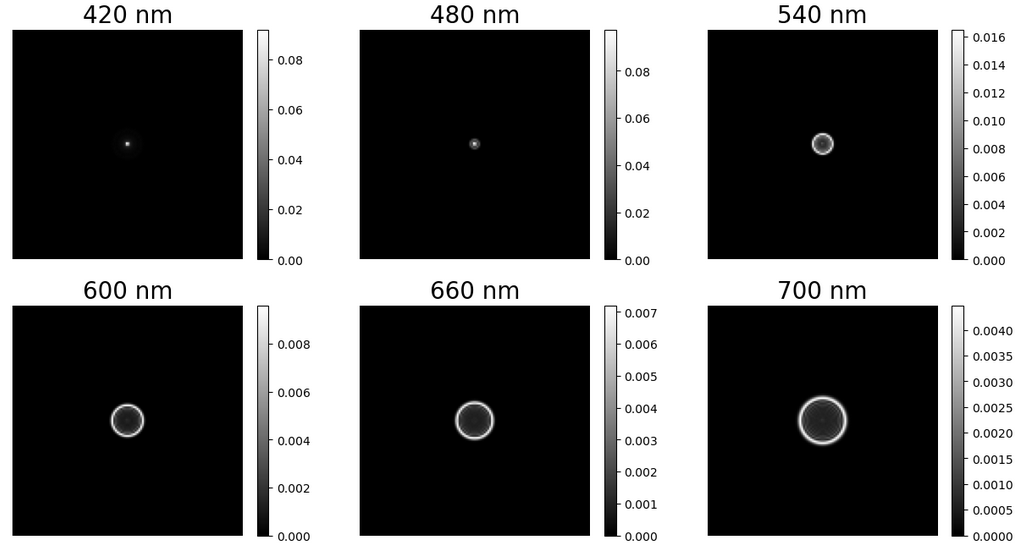}
    \caption{PSFs of 6 wavelengths at one location (zoomed in 4 times).}
    \label{PSFs}
\end{figure}

\begin{figure}[h]
    \centering
    \includegraphics[width=0.9\linewidth]{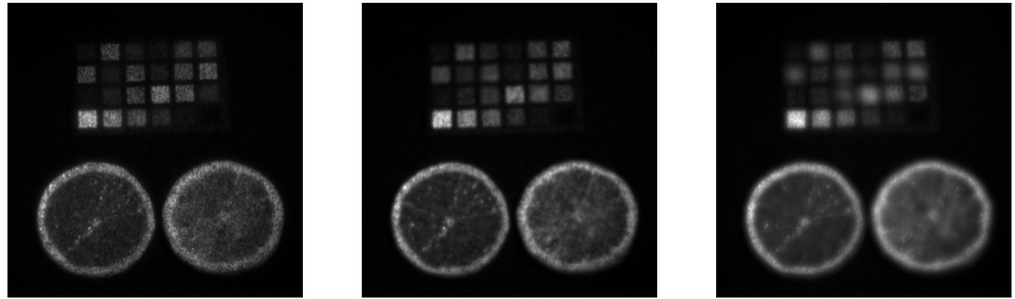}
    \caption{Simulation of three measurements on the detector at three different locations and with different coding masks.}
    \label{Measurements}
\end{figure}

\begin{figure}[h]%
    \centering
    \subfloat[Ground Truth (RGB).]{
    \includegraphics[width=0.35\linewidth]{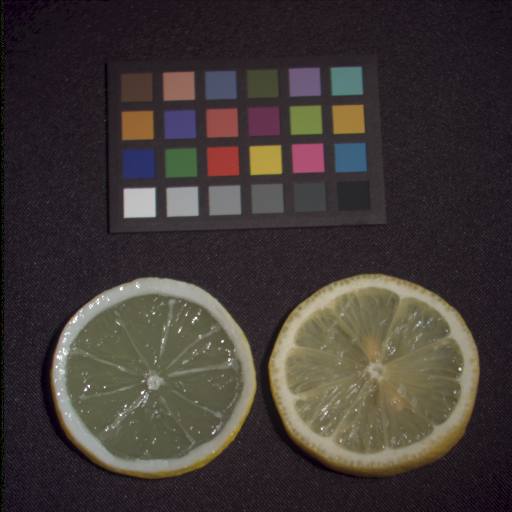}
    \label{OG}
    }
    \subfloat[RGB reconstruction (PSNR = 35.46 dB).]{
    \includegraphics[width=0.35\linewidth]{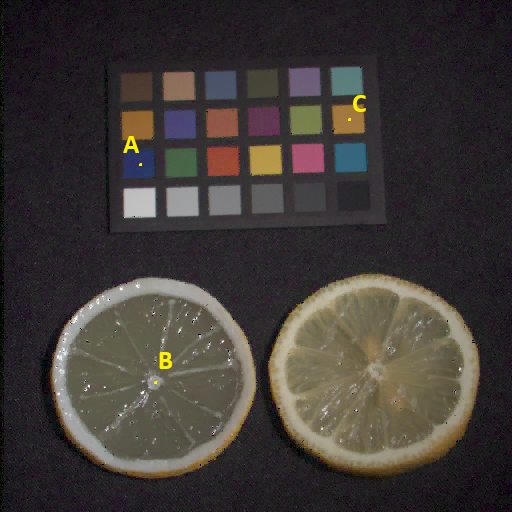}
    \label{rec_rgb}
    }\\
    \subfloat[Spectral reconstruction of 3 points.]{
    \includegraphics[width=0.65\linewidth]{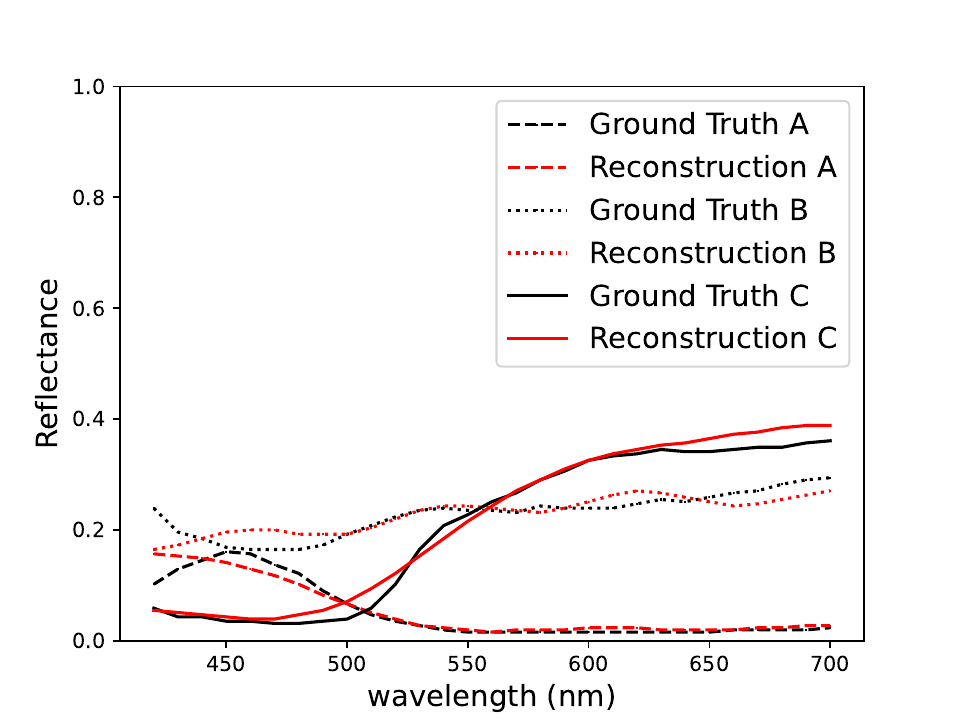}
    \label{rec_spec}
    }
    \caption{Image reconstruction. Since three measurements are used for the reconstruction, and there are 29 spectral bands to retrieve, the compression ratio is roughly 10\%.}
\end{figure}

Simulations are performed to test the feasibility of the method. The crucial task is extracting PSFs from a refractive lens to confirm it exhibits adequate dispersion. To do so, a lens is selected in a commercial catalog to present the desired chromatic aberration. This lens is introduced in the \textit{Code V} software. The latter enables retrieval of PSFs at specific wavelengths and different detector locations. An example of 6 PSFs at different wavelengths across the spectrum and at one location are shown in Fig. \ref{PSFs}.\\
A hyperspectral image from the \textit{CAVE} \cite{CAVE_0293} dataset is used as test data. It can be seen in Fig. \ref{OG} and contains 29 spectral bands ranging from 420 to 700 nm by 10 nm steps. Hence, three sets of 29 PSFs at the required wavelengths are taken at three different detector locations. To simulate a measurement, all spectral bands are first coded with the same random binary mask, then each coded band is convolved with its corresponding PSF. Finally, all the resulting bands are summed, which yields the measurement. The three measurements from which the following reconstruction is performed are shown in Fig. \ref{Measurements}.

Afterwards, these measurements are injected into the reconstruction algorithm along with their corresponding coding masks and PSFs. The RGB representation of the reconstruction is shown in Fig. \ref{rec_rgb}, along with the spectral reconstruction of 3 points in Fig. \ref{rec_spec}.

\section{Conclusion}
Simulations show promising results for the design of a compact compressive spectral imager suitable for Earth observation that requires only a collecting lens, a SLM, a dispersive lens, and a monochrome detector. Future work will evaluate our approach on  measurements obtained in laboratory from an actual setup, using a robust estimate of their noise level $\epsilon$ for the image estimation.

\bibliographystyle{IEEEtran}
\bibliography{references}
\end{document}